\documentclass{rspublic}
\usepackage{amsmath}
\usepackage{amsfonts}
\usepackage{amssymb}
\usepackage[Symbol]{upgreek}
\usepackage[nointegrals]{wasysym}
\usepackage{mathrsfs}
\usepackage{graphicx}
\newcommand{\real}{{\mathbb R}}
\newcommand{\bfu}{{\bf u}}
\newcommand{\bfx}{{\bf x}}
\begin{document}
\title[Navier-Stokes Regularity]{\large Remarks on backward uniqueness of parabolic equations and incompressible Navier-Stokes well-posedness}
\author[F. Lam]{F. Lam}
\label{firstpage}
\maketitle
\begin{abstract}{Navier-Stokes Equations; Incompressibility; Backward Uniqueness; Carleman Inequality; Parabolic Scalings; Regularity}

We explain why the theory of Escauriaza, Seregin, and \v{S}ver\'{a}k (Russian Math. Surveys, 2003) on potential finite time singularity in Navier-Stokes solutions must be largely misapprehended. It is found that the proofs of the backward uniqueness theorem for parabolic equations contain technical errors. The stated validity of a theorem for vorticity is established on ill-informed analyses as the solenoidal constraint is not taken into account. There are many cases where parabolic scalings are erroneously applied. We briefly discuss a number of related issues.
\end{abstract}
\section{Background}
In the Eulerian description of the motion of an incompressible, homogeneous Newtonian fluid, the momentum, and the continuity equations are 
\begin{equation} \label{ns}
	{\partial {\bfu}}/{\partial t} + ({\bfu} . \nabla) {\bfu} = \nu \Delta {\bfu}   - {\rho}^{-1} \nabla p, \;\;\; \nabla . {\bfu} =0,
\end{equation}
where the velocity vector ${\bfu}={\bfu}({\bfx},t)$ has the components $(\bfu_1,\bfu_2,\bfu_3)$, the scalar quantity $p=p({\bfx},t)$ is the pressure, the space variable is denoted by ${\bfx}=(\bfx_1,\bfx_2,\bfx_3)$, and $\Delta$ is the Laplacian. The kinematic viscosity is $\nu =\mu/\rho$, where $\rho$ and $\mu$ are the density and viscosity respectively. These equations are known as the Navier-Stokes equations. 

We seek the solution of (\ref{ns}) as an initial-boundary value problem in the space-time domain denoted by $\Upomega_0 {\times} [\:0,t {<} T]$, where $0{<}T{<}\infty$. In the present study, $\Upomega_0$ denotes the whole space $\real^3$ or the half-space $\real^3_+$. 
We are interested in flow evolution from given data of finite energy. The initial condition is given by
\begin{equation} \label{ic}
 {\bfu}({\bfx},t{=}0) = {\bfu}_0({\bfx}) \; \in \; C_c^{\infty}(\Upomega_0),
\end{equation} 
and $\nabla.\bfu_0=0$, in the case of the half-space, the no-slip boundary condition
\begin{equation} \label{bc}
 {\bfu}({\bfx},t) = 0\;\;\; \forall z = 0.
\end{equation} 

Taking divergence of the momentum equation in (\ref{ns}), and making use of the continuity, we obtain a Poisson equation for the pressure 
\begin{equation} \label{pp}
	\Delta p(\bfx)= - \rho \sum^{3}_{i,j=1} \Big( \frac{\partial \bfu_j}{\partial \bfx_i} \: \frac{\partial \bfu_i}{\partial \bfx_j}\Big)(\bfx).
\end{equation}
We would like to emphasise the fact that, at every instant of time, equation (\ref{pp}) {\it does not} describe the dynamic evolution of the pressure as it contains no time information. 

A physical (vector) quantity known as vorticity, $\omega=\nabla \times \bfu$, is related to the dynamics of angular momentum in flow motions. In view of the vector identity $\nabla{\times}(\nabla A)=0$, the concept of vorticity enables us to consider flow motions with three unknowns instead of four. Vorticity evolution is governed by the vorticity equation
\begin{equation} \label{vort}
	{\partial \omega}/{\partial t} - \nu \Delta \omega = (\omega . \nabla) \bfu  - (\bfu . \nabla )\omega.
\end{equation}
The vorticity field inherits the velocity solenoidal property of (\ref{ns})
\begin{equation} \label{incp}
\nabla. \omega=0. 
\end{equation}
This is a direct consequence of the vector identity $\nabla.(\nabla{\times}A)=0$. For the Cauchy problem, the initial vorticity is specified as 
\begin{equation} \label{vic}
 \omega (\bfx,0) = \nabla {\times} \bfu_0 (\bfx) = \omega_0 (\bfx) \; \in \; C_c^{\infty}(\Upomega_0).
\end{equation} 
In the case of the half-space, the value of vorticity on $z=0$ is unknown {\it a priori} and must be determined as part of the solution.

The velocity can be recovered from the Biot-Savart relation
\begin{equation} \label{bs}
\bfu(\bfx)= \frac{1}{4 \pi} \int \frac{(\bfx-\bfx')}{|\bfx-\bfx'|^3} {\times} \omega(\bfx') \rd \bfx'.	
\end{equation}
It is instructive to notice that this relation as well as the pressure given by (\ref{pp}) illustrate the {\it mutual global dependence} of the flow variables at every instant of time. Under the hypothesis of incompressibility, any space-time perturbation in the vorticity will instantaneously propagate to every part of $\real^3$ and thus modify the velocity (and the pressure) which in turn may generate vorticity gradients. Hence a local energy, the energy of fluid over a finite space, can only exist momentarily in flows of bounded energy, except in vorticity-free motions. Nevertheless, the dynamic properties of the velocity and the pressure can be derived from the momentum equation in (\ref{ns}) or from its time derivatives.

By observation, governing equations (\ref{ns}) and (\ref{vort}) are invariant under time translation $t \rightarrow t + t_s$ for finite constant $t_s$ which, clearly, refers to a starting instant. The time symmetry is just Noether's invariance theorem for energy conservation. Because the equations describe an irreversible process of diffusion in accordance with the second law of thermodynamics, fluid dynamics in backward time $[-T_0,0], 0{<}T_0{<}\infty$ must be investigated in terms of time-translated equations of motion over interval $[0,T_0]$. Fundamentally, evaluations of time-shifted solutions, either weak or ancient, do not shed new light on regularity.

Given a {\it positive} parameter $\lambda$, one can perform algebraic manipulations of (\ref{ns}) and (\ref{vort}) by means of transformations $\bfx \rightarrow (\bfx/\lambda)$ and $t \rightarrow (t/ \lambda^2)$. The dimensional analysis shows that if a triplet $(\bfu,p, \omega)$ solves the Navier-Stokes system, so does $(\bfu_{\lambda},p_{\lambda}, \omega_{\lambda})$, where
\begin{equation} \label{scaling}
  \begin{split}
	\bfu_{\lambda}( \bfx,t) & = {\lambda}^{-1} \bfu\: ( {\lambda}^{-1} \bfx, {\lambda}^{-2} t ), \\
	p_{\lambda}( \bfx,t) & = {\lambda}^{-2} p\:( {\lambda}^{-1} \bfx, {\lambda}^{-2} t), \\
	\omega_{\lambda} (\bfx,t) & = {\lambda}^{-2} \omega \: ( {\lambda}^{-1} \bfx, {\lambda}^{-2} t), \\
	\end{split}
\end{equation}
as fluid properties $\rho$ and $\mu$ are assumed homogeneous. Similar scalings apply to the heat equation $\partial_t Q - \kappa \Delta Q = 0$, where $Q$ may be interpreted as temperature. Thus $Q_{\lambda}(\bfx,t)=Q( {\lambda}^{-1} \bfx, {\lambda}^{-2} t )$.
The parabolic scalings are relevant in transforming one set of {\it continuous} triplet solution into a dilated one. The space-time dilation or contraction is limited by the fundamental dimensions of matters; the scaling parameter $\lambda$ cannot be arbitrarily assigned. Formally, triplet $(\bfu_{\lambda},p_{\lambda}, \omega_{\lambda}) \rightarrow 0$ as $\lambda \rightarrow \infty$ for bounded solution $(\bfu, p, \omega)$. The limit is consistent with the continuum hypothesis. 
If a triplet $(\tilde{\bfu}, \tilde{p}, \tilde{\omega})$ is shown to exist only in a space of equivalence classes of functions, for instance $L^q(\Upomega_0), 1\leq q < \infty$, it makes no sense to apply the scalings to this triplet at any specific space-time location. 

Attempts have been made in the case of invariant norm $\|\bfu\|_{L^3(\real^3)L^{\infty}[0,T_s)}$.  Specifically, Escauriaza {\it et al.} (2003$b$) claimed that the local-in-time regularity under the invariant scalings could be settled by means of a backward uniqueness of parabolic equations in conjunction with certain Carleman inequalities. The regularity interval is denoted by $[0,T_s)$, where $0<T_s<\infty$. Should a singularity develop at time $T_s$ from initial data (\ref{ic}), the solution would blow up at all of the possible flow scales or
\begin{equation} \label{b1}
	\lim_{t\rightarrow T_s} \sup \| {\bfu}(\cdot,t) \|_{L^3(\real^3)} \rightarrow \infty.
\end{equation}

Our priority is to draw our attention to the analytical methods involved, as any new techniques of attacking the regularity problem ought to attract wider interest. 
\section{Backward uniqueness for parabolic equations}
Escauriaza {\it et al.} (2003$a$) intended to show a backward uniqueness theorem for the solution of $\mathscr{D}u=0$, where the parabolic operator
\begin{equation} \label{po}
	\mathscr{D}=\partial_t - \Delta + b(x,t) \nabla_x + c(x,t). 
\end{equation}
It is assumed that the solution $u$ is bounded and decays at infinity. The functions $b$ and $c$ are measurable and bounded, satisfying certain regularity and boundedness constraints. Denote $\Upomega_B = {\real^n \backslash B_R}$ , where the hyper-sphere $B_R=\{x \in \real^n: |x| \leq R \}$. Suppose that $u=u(x,t)$ is a bounded solution of $\mathscr{D}u=0$ in the parabolic domain $\Upomega_T = \Upomega_B \times [0,T>0]$. If $u(\cdot,T) = 0 $ in $\Upomega_B$, then $u \equiv 0$ in $\Upomega_T$. Of course we are mainly interested in the case of $n=3$.

The proof of the backward uniqueness depends crucially on an inequality of Carleman type, i.e. equation (1.5) of Escauriaza {\it et al.} (2003$a$). For definiteness, we write it down in full: 
{\it There is a constant $\alpha_0=\alpha_0(R,n)$ such that the inequalities
\begin{equation} \label{c2}
\begin{split}
	\| \re^Z u\|_{{L^2}(\Upomega_T)} + \|\re^Z \nabla u & \|_{{L^2}_(\Upomega_T)} \\
	& \leq \|\re^Z (\partial_t u + \Delta u)\|_{{L^2}(\Upomega_T)} + \|\re^{|x|^2} \nabla u (\cdot,T)\|_{{L^2}(\Upomega)}
\end{split}
\end{equation}
hold for all $\alpha \geq \alpha_0$ and $u \in C_c^{\infty}(\Upomega_T)$ satisfying $u(\cdot,0)=0$, and in the exponential weight function $Z=\alpha(T-t)(|x|-R)+|x|^2$}. 

On pages 153-155 for the proof of Lemma 4, the last displayed equation reads:
\begin{equation*} 
	\|u\|_{L^2(\Upomega')} \lesssim \exp\Big( \alpha r - r^2 \Big) + \exp\Big( - \alpha a \varepsilon R \Big),
\end{equation*}
where the domain $\Upomega' = B_r \backslash B_{(1+10a)R} \times [0,\varepsilon>0]$. The time interval is sufficiently small. It was claimed that $u \equiv 0$ in $\Upomega_B {\times} [0,\varepsilon]$ in the limit of first $r \rightarrow + \infty$, then $\alpha \rightarrow + \infty$ and finally $a \rightarrow 0^+$.  

In fact, the limit is ambiguous because 
\begin{equation*}
	\lim_{r \rightarrow +\infty} \exp\Big( r \big( \alpha - r\big) \Big) \longrightarrow \exp\Big((+\infty) \times (- \infty )\Big)\;\;\; \mbox{or} \;\;\; \lim_{r \rightarrow +\infty} \frac{\exp ( \alpha r)} {\exp ( r^2 )} \longrightarrow \frac{\infty}{\infty}
\end{equation*}
for finite positive $\alpha$. If we let both $r \rightarrow + \infty$ {\it and } $\alpha \rightarrow + \infty$ as these two parameters are independent of each other, we obtain 
	$\lim_{r, \alpha \rightarrow +\infty} \; \rightarrow \; \exp (\infty - \infty)$. 
All of these limits are indeterminable in the analysis of the extended reals.

Furthermore, suppose that $r$ remains finite while the parameter $\alpha$ is allowed to increase arbitrarily to infinity. Evidently, the second exponential function on the right vanishes for finite positive $a$ and $\varepsilon$ while the first exponential tends to $+\infty$.

There are similar technical inconsistencies in the decay limit on the last term on the right-hand side of equation (2.11). The term is bounded by
\begin{equation*}
	a^{-k} \; \exp(- r^2) \; \|u\|_{L^{\infty}(B_{4r}{\backslash}B_{2r}){\times}[0,1]}\;\;\; k\geq0.
\end{equation*}
This bound tends to zero as the exponential function decays in the limit of $r \rightarrow \infty$, {\it only if $a$ is non-zero for} $k \geq1 $ so as to avoid indeterminacy $\infty \times 0$. However, the space-time parabolic cylinder
\begin{equation*}
	\lim_{r \rightarrow \infty}\; (B_{4r}{\backslash}B_{r}){\times}[0,1] \;\longrightarrow \;\varnothing.
\end{equation*}

Because the limit for $\|u\|_{L^2}$ does not exist as $r {\rightarrow} \infty$ regardless of whether parameter $\alpha $ tends to $\infty$ or remains finite, solution $u$ in Lemma 4 does not necessarily vanish in the parabolic cylinder. We assert that {\it the proof of the backward uniqueness for parabolic equations (Theorem 1) in the punched cylinder is incomplete and unjustified}.
\section{Weight function in anisotropic Carleman inequality}
To apply the backward uniqueness to the Navier-Stokes equations, a Carleman inequality, with a revised weight function compared to (\ref{c2}) above, is introduced in ``the half-space'' $\real^n_+$. Following Escauriaza {\it et al.} (2003$b$), we state their proposition 6.2 or equation (6.12) as follows:
{\it Let 
\begin{equation*}
	\phi(x,t) = \phi^{(1)}(x,t) + \phi^{(2)}(x,t),
\end{equation*}
where
\begin{equation*}
	\phi^{(1)}(x,t) = - \frac{|x'|^2}{8 t}\;\;\;, \phi^{(2)}(x,t) = a(1-t) \frac{x_n^{2 \alpha}}{t^{\alpha}},
\end{equation*}
$x'=(x_1,x_2, \cdots , x_{n-1})$ so that $x=(x',x_n)$, and $e_n=(0,0,\cdots,0,1)$. Then, for any compactly-supported function
\begin{equation*}
	u \in C_c^{\infty} ( \Upomega_1; \real^n),
\end{equation*}
where $\Upomega_1=(\real^n_+ {+} e_n) {\times} (0,1)$, and for \underline{any number $a> a_0(\alpha)$}, the following inequality is valid:
\begin{equation*}
\begin{split}
	\int_{\Upomega_1} t^2 \exp\big(2 \phi\big)  \big(a |u|^2/t^2+ &|\nabla u|/t \big) \rd x \rd t \\ & \leq c_{\star}\int_{\Upomega_1} t^2 \exp\big(2 \phi\big) | \partial_t u + \Delta u |^2 \rd x \rd t.
\end{split}
\end{equation*}
Here $c_{\star}=c_{\star}(\alpha)$ is a positive constant and $\alpha \in (1/2,1)$ is fixed.}

The symbol $\alpha$ here should not be confused with the one as discussed in the previous section. We emphasise the fact that the specific nature of the weight function is essential; the entire function $\exp(2 \phi)$ in the above inequality must be bounded in order for the Carleman inequality to be meaningful. In their proof of the proposition, function $A_2$ was defined by
\begin{equation*}
	A_2 = - \partial_t|\nabla \phi^{(2)}|^2 - \Delta ^2 \phi^{(2)} - |\nabla \phi^{(2)}|^2 /t,
\end{equation*}
and, it was shown that
\begin{equation*}
	A_2 \geq a \:\frac{1-t}{t^{\alpha}} x_n^{2 \alpha - 4} (2\alpha -1)\; \Big( a \: \frac{4 \alpha^2 x_n^{2 \alpha+2}}{t^{\alpha+1}} - 2 \alpha(2 \alpha-2)(2 \alpha -3)\Big) \equiv K > 0.
\end{equation*}

Since $A_2$ depends on the weight $\phi^{(2)}$ and hence on the number $a$ (in fact it is a free positive parameter), then the following inequalities must hold
\begin{equation*} 
	A_2 +  \Big( \frac{a}{a-1} \Big)^a > K > 0
\end{equation*}
for all $a > 2$. As $a \rightarrow + \infty$, the left-hand reduces to $A_2 + 1^{\infty}$ which is an indeterminate form. Consequently, integral $I$ in Proposition (6.1) does not exist in the limit. The contradiction suggests that {\it the (arbitrary) parameter $a$ in the Carleman inequality must be finite} so that, at least, the weight $\phi$ needs to be bounded in any non empty open set $\Upomega_+ \subset \real^n_+$ while time $t$ is less than unity. 

One of the implications of $a$ finiteness is that the proof of Lemma 5.3 is technically flawed. The last displayed equation in the proof is 
\begin{equation*} 
\begin{split}
	\int_{\Sigma} \exp(2 a \phi_B) \big(|u|^2 {+} |\nabla u|^2\big)(\lambda y, 
	\lambda^2(s {-}1/2)) &(s y_n)^2 \exp\Big({-}\frac{|y'|^2}{4s}\Big) \chi\rd y \rd s \\
	& \leq d \exp\Big( - 2^{\alpha-1} \Big(\frac{3}{\lambda}+2\Big)_n^{2 \alpha} \; a\Big)
	\end{split}
\end{equation*}
where $d=d(\lambda, c_{\star})$ is a bounded function, and $\lambda \in (0,1/\sqrt{12}\:]$, integration domain $\Sigma$ denotes the parabolic cylinder $(\real^n_++(3/\lambda{+}1)e_n)\times (1/2,1)$, and  
the cut-off function 
\begin{equation*}
	\chi=\chi(y_n,s)=\begin{cases}
	1 & \text{if} \;\;\; (y_n,s) \in W,\\
	0 & \text{if} \;\;\; (y_n,s) \notin W,
	\end{cases}
\end{equation*}
and 
\begin{equation*}
	W \equiv \Big\{ (y_n,s) \; \| \;\; y_n>1,\; 1/2<s< 1, \;\; a \phi_B(y_n,s) < - \phi^{(2)} \Big( \frac{3}{\lambda}+2,\frac{1}{2}\Big)\Big\}.
\end{equation*}
It was asserted that $u(x,t)=0$ for all $x \in \real^n_+$ and for all $t \in (0,1/24)$ because, in the limit $a \rightarrow +\infty$ or effectively $\phi^{(2)} \rightarrow +\infty$, the exponential function rendered the right-hand side to zero. We believe that this operation of taking the limit is misleading; we have just shown that the Carleman inequality (6.12) has no meaning when $\phi \rightarrow \infty$. On the other hand, it is not clear how the analyses in the proofs of Lemmas 5.2-5.4 may determine the required vanishing condition $u(x,t)\equiv0$ for any bounded $a$. Certainly, {\it the backward uniqueness for the heat equation in the half-space as stated in Theorem 5.1 is unattainable for unbounded weight $\phi$}.
\section{Incompressibility and parabolic scalings}
In the problem of Navier-Stokes regularity in three-space dimensions, it is vorticity equation (\ref{vort}) which may be considered as a heat equation as $\mathscr{D}\omega=0$. As the vorticity field is solenoidal (cf. (\ref{incp})), the parabolic solution in any backward uniqueness, and any Carleman inequality, (the function denoted by $u$) must be divergence-free. 

In the proof of proposition 6.1 of Escauriaza {\it et al.} (2003$b$), the scaled function $v$ is defined by $v(x,t)=(\exp(\tilde{\phi})u)(x,t)$. The weight depends on space and time:   
\begin{equation*}
	\tilde{\phi}(x,t) = - {|x|^2}/{(8t)} - (a+1) \log(h(t)), 
\end{equation*}
where $a>0$ is a real number, and the time-dependent function $h=t\exp((1-t)/3)$ for $t \in (0,2)$. Clearly $\nabla.\tilde{\phi}\neq0$ unless $x\equiv0$ and thus neither $u$ nor $v$ can be solenoidal. (We have used the tilde symbol to distinguish the weight from the one defined in proposition 6.2.)
By the same token, the weight function of proposition 6.2 as written above $\nabla.\phi(x)\neq0$ except at the space origin for every instant $t$. As a result, the scaled function $v(x,t)=(\exp(\phi)u)(x,t)$ used throughout equations (6.2)-(6.6) is {\it not} solenoidal, because a solution $u$ of the parabolic equation $\mathscr{D}u=0$ is not necessarily divergence-free under bound (5.1) and decay (5.3). The only exception is the trivial solution $\omega = u = v = 0$. Our discussion implies that {\it the conclusions drawn from the formal application of the backward uniqueness Theorem 5.1 to the Navier-Stokes vorticity, as stated in their \S3, are simply false}.

It was stated that the pressure was bounded (cf. (3.4))
\begin{equation*}
	p \in L^{3/2}(\real^3)\;L^{\infty}(0,T).
\end{equation*}
This is incorrect. The {\it a priori} bound obtainable by {\it time} integration of the pressure solution from (3.3) is no more than the summability:
\begin{equation*}
	p \in L^{3/2}(\real^3)\;L^{1}(0,T).
\end{equation*}
Evidently, the bound of Caffarelli-Kohn-Nirenberg, $p \in L^{5/3}(\real^3)\:L^{5/4}(0,T)$, is an improvement (Caffarelli {\it et al}. 1982).
In view of time-wise embedding ($1 {<} 5/4 {<} 3/2$), {\it the pressure does not possess all of the regularity as required by Theorem 1.4}. 

There are a number of occasions where the meaning of the space-time norms (denoted by $L_{s,l}(Q_T)$) causes confusion. To be specific, consider the velocity in \S 3 (cf. (3.2)): $v \in L^{4}(\real^3)\;L^4(0,T)$. In order to show H\"{o}lder continuity, $v$ was scaled: ${\tilde v}(x,t)=Rv(\tilde x, \tilde t)$, where $\tilde x = x_0+Rx$, and $\tilde t=t_0+R^2t$ ($R{>}0$). Evidently, the velocity was treated as functions. So let us consider 
\begin{equation*}
{\tilde v}(x,t) \sim {\tilde t}\: {\tilde x}^{-3/4+\varepsilon}, \; \varepsilon >0.
\end{equation*}
Clearly, the $L^4$-norm of $\tilde v$ exists over $B_1{\times}(0,1)$ but $v$ is unbounded at $x=-R/x_0$. Recall the fact that regularity of the weak solutions has not yet been proven at this stage of the analysis. In general, if $v \in L^q(B_1{\times}(0,1]),\: 1\leq q < \infty$, the singularity of counter example, $v(x,t) \sim (x\:t)^{(\varepsilon-3)/q}$ at the space origin, demonstrates the fact that {\it exercising parabolic scalings (\ref{scaling}) to the weak solutions is unjustifiable}. Regrettably, {\it the proof of Theorem 1.3 is misguided}.

In theory of parabolic equations, fundamental solutions for operator (\ref{po}) are established only for time $t>0$ due to irreversibility (for adjoint in reversed time). In homogeneous media filling $\real^3$, the fundamental solution does not suggest an emergency stop mechanism for viscous diffusion, which is in essence energy-dissipative on fine-structures in unrestrained space over time-scales substantially longer than macroscopic dynamics. The specification of initial data (\ref{ic}), such as size and decay, is paramount in initial value problem (\ref{vort}). Guided by physics and experiment, it is reasonable to envisage a reality where a proposition of backward uniqueness, independent of problem's initial (and boundary) conditions, would not exist in all likelihood. Basically, {\it dilating Navier-Stokes solutions into negative time zone by parabolic scalings (\ref{scaling}) violates the second law of thermodynamics}. (A scaled solution that disrespects the law may be bounded.)  
\section{Outlook}
We find that the proofs for the local regularity and potential blow-up stated by Escauriaza, Seregin, and \v{S}ver\'{a}k (2003$a$; 2003$b$) contain analytical mistakes as well as misconceptions of serious nature. Claims and applications made on the basis of the proposed problem-solving scheme require critical reappraisals. 

Any incompressible fluid motion involves velocity, pressure, as well as vorticity; the aggregate flow field is scale-sensitive in norm $L^{3}_x \: L^{\infty}_t$. Equally, we may as well consider the vorticity equation in $L^{3/2}_x \: L^{\infty}_t$. There does not exist one single definitive scale in which triplet $(\bfu, p, \omega)$ is invariant during flow evolution. On the other hand, the energy identity of regular Navier-Stokes solutions, 
\begin{equation*} 
	\frac{1}{2}\;\big\|\bfu(t)\big\|^2_{L^2 (\real^3)} + \nu \int_0^T \! \big\|\nabla \bfu (t) \big\|^2_{L^2 (\real^3)} \rd t  \; = \; \frac{1}{2}\;\big\|\bfu_0 \big\|^2_{L^2 (\real^3)},
\end{equation*}
is truly scale-independent. Both the velocity and its Jacobian are taken into account, and, implicitly, the pressure or the continuity is also effected. The conservation law does not demand any scale-dependency on initial data (\ref{ic}). By virtue of the Biot-Savart relation (\ref{bs}) or the pressure gradient from (\ref{pp}), triplet $(\bfu, p, \omega)$ is essentially non-local. The solutions are known to be globally regular; no singularity is possible on any local scales. Experience shows that turbulent motions decay in time. The last stage of the decaying flow may be modelled approximately by diffusion equation, $\partial_t \bfu - \nu \Delta \bfu {=}0$, that has only constant-scale solutions. Thus parabolic scalings (\ref{scaling}) cannot play a significant role in understanding the spatio-temporal structures of turbulence that are theoretically tractable by a proper treatment of the non-linearity. {\it The finite-time singularity scenario (\ref{b1}) is illusive}.
%
%
\addcontentsline{toc}{section}{\noindent{References}}

\begin{acknowledgements} 
\vspace{0.5mm}

\noindent 
16 April 2019

\noindent 
\texttt{f.lam11@yahoo.com}

\end{acknowledgements}
%
\end{document}